\documentclass[twocolumn,showpacs,showkeys,preprintnumbers,amsmath,amssymb,superscriptaddress,prl]{revtex4}

\usepackage{graphicx}
\usepackage{dcolumn}
\usepackage{bm}
\usepackage[latin1]{inputenc}

\begin{document}
\title{Metal-insulator transition in one-dimensional lattices with chaotic energy sequences}
\author{R. A. Pinto}
\affiliation{Laboratorio de F\'\i sica Estad\'\i stica, Centro de F\'{\i}sica, Instituto Venezolano de Investigaciones Cient\'{\i}ficas, Apartado 21827, Caracas 1020-A,
Venezuela
}
\author{M. Rodr\'{\i}guez}
\affiliation{Laboratorio de F\'\i sica Estad\'\i stica, Centro de F\'{\i}sica, Instituto Venezolano de Investigaciones Cient\'{\i}ficas, Apartado 21827, Caracas 1020-A,
Venezuela
}
\author{J. A. Gonz\'alez}
\affiliation{Laboratorio de F\'\i sica Computacional, Centro de F\'{\i}sica, Instituto Venezolano de Investigaciones Cient\'{\i}ficas, Apartado 21827, Caracas 1020-A,
Venezuela
}
\author{E. Medina}
\affiliation{Laboratorio de F\'\i sica Estad\'\i stica, Centro de F\'{\i}sica, Instituto Venezolano de Investigaciones Cient\'{\i}ficas, Apartado 21827, Caracas 1020-A,
Venezuela
}

\date{\today}

\begin{abstract}
\hspace{0.27cm} We study electronic transport through a one-dimensional array of sites by using a tight binding Hamiltonian, whose site-energies are drawn from an chaotic sequence . The correlation degree between these energies is controlled by a parameter regulating the dynamic Lyapunov exponent measuring the degree of chaos. We observe the effect of chaotic sequences on the localization length, conductance , conductance distribution and wave function, finding evidence of a Metal-Insulator Transition (MIT) at a critical degree of chaos.  The one dimensional metallic phase is characterized by a Gaussian conductance distribution and exhibits a peculiar non-selfaveraging.
\end{abstract}

\maketitle

It is well-known that one-dimensional systems with uncorrelated disorder behave like insulators because their electronic states tend to localize\cite{Anderson}. For systems whose length is larger than the electronic localization length the conductance vanishes exponentially.  Nevertheless, the Anderson localization\cite{AALR} does not address the effect that correlations have on electronic states in disordered lattices. Short range correlations have been shown to be responsible for low dimensional delocalization behavior in direct experimental realizations using semiconductor superlattices\cite{dimer1,dimer2,dimer3}. Such is the behavior of  the one-dimensional random-dimer model, where transitions from localized to extended states are found when the Fermi energy assumes values belonging to a discrete set of energies\cite{dimer4}. 

Extended states have been reported for incommensurate one dimensional systems which can be viewed as long range correlated. This is known to occur within the Harper model\cite{Hiramoto} whose the site energies are given by $\epsilon_n=(w/2)\cos(2\pi n\omega)$ where $\omega$ is irrational and $w$ is a parameter. Below a critical value of $w$ extended states are obtained. A more general model was introduced by Griniasty and Fishman\cite{Griniasty} including the Harper model as a particular case,  describes a one dimensional model where the energy $\epsilon_n$ of site $n$ is generated by the relation $\epsilon_n=\cos(\pi \alpha n^\nu)$. The resulting sequence is described as pseudorandom depending on the value of $\nu$ and $\alpha$ an irrational number. In this model a mobility edge is found whereupon the system exhibits a metal insulator transition for a threshold value of $\nu$.

Mobility edges in one dimensional systems have also been recently reported in non-periodic Kronig-Penney potentials correlated over long distances\cite{Izrailev}. Here it was shown how to construct site energy sequences, that lead to predetermined localization-delocalization transitions. Interestingly, mobility edges have been found experimentally for these systems by microwave transition measurements on waveguides with inserted correlated scatterers\cite{Izrailev2}.  More recently, non-trivial electronic transport properties in low dimensional systems such as certain proteins, and DNA chains\cite{Flores, dimer1} have prompted new models involving long range correlated disorder\cite{Lyra}. 
 
In this work, we study electronic transport through a quasi-one dimensional system
with site energies obeying a chaotic sequence characterized by a dynamic Lyapunov exponent $\lambda$, that gives a measure of the degree of chaos and {\it short ranged correlation}. The model of chaotic sequences used here is of much interest since it addresses in a continuos fashion periodic and chaotic sequences of varying dynamic Lyapunov exponent with a single control parameter $z$. The latter parameter can in turn be related analytically to the dynamic Lyapunov exponent, the rate at which information is lost in the sequence. Random sequences are then identified with large values of the dynamic Lyapunov exponent were no memory of the previous values is preserved\cite{random}, while small values of the exponent relate to correlated sequences. The manner of introducing disorder is reminiscent of that proposed in ref. \cite{Griniasty} discussed above, where a parameter served to measure the degree of pseudo-randomness. Pseudo-randomness, in this case, is well defined by a sequence of numerical tests in ref. \cite{Brenner} in relation to Anderson localization. 

This work first addresses the chaotic energy sequence recursion and the analytic form of the dynamic Lyapunov exponent. The methods for obtaining reliably the wave function and the localization length as a function of energy follow, along with the computation of the conductance of the system from the Green's function. We then show that the localization length grows faster than the length of the system, within a finite energy band, below a critical value of the parameter governing the degree of chaos $z$, indicating delocalized electron behavior. Such a conclusion is supported by studying the conductance as a function of system size, its asymptotic scaling properties, and the localization of  the wavefunction. The conductance distribution in the extended phase is a narrow Gaussian for different realization of the chaotic sequence but its mean fluctuates more strongly with system size in comparison to the distribution width, displaying non-self averaging behavior.

We shall consider a tight binding Hamiltonian with nearest-neighbor coupling in order to describe one electron moving through a chain of $N$ sites
\begin{equation}\label{eq:TB}
H = \sum_{n} \epsilon_n |n\rangle\langle n| + \sum_{\langle n,m\rangle}
V|n\rangle\langle m| \; .
\end{equation}
where $\epsilon_n$ are the site energies and $V$ the nearest neighbor coupling. This chain is attached to semi-infinite leads at each end, whose site energies are equal to $\epsilon_0$, and the hoppings are unity. The site energies in the sample are generated by the recursive formula (inspired by the recursion of Nazareno, Gonzalez and Costa (NGC)\cite{Nazareno} )
\begin{equation}\label{eq:recursion}
\epsilon_{n+1} = 1 - \sin^2[z \arcsin{\sqrt{\epsilon_n}}] \; ,
\end{equation}
$z$ being the controlling parameter that regulates the dynamic Lyapunov exponent. By using the transformation $y_n = \frac{2}{\pi}\arcsin{\sqrt{\epsilon_n}}$, we obtain a piecewise linear map $y_{n+1}=f(y_n)$. One can then find an analytic expression for the dynamic Lyapunov exponent by applying the relation
\begin{equation}
\lambda = \lim_{N\to\infty} \frac{1}{N} \sum \ln \left|\frac{df(y_n)}{dy_n} \right|
=\ln z.
\end{equation}
Therefore, for $z>1$ the map is chaotic\cite{Nazareno}.

\begin{figure}
\includegraphics[width=3in]{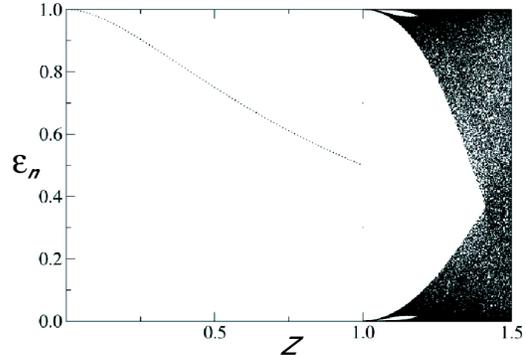}
\caption{\label{figure1}Bifurcation map corresponding to the recursive relation of Eq.
\ref{eq:recursion}. The sequence at the threshold oscillates between essentially two values (0 and 1), but gradually grows to the full interval (0,1).  The sequence is only strictly random in the $z>>1$ limit\cite{Nazareno}.}
\end{figure}

In Fig.~\ref{figure1} we show a bifurcation map corresponding to the formula (\ref{eq:recursion}). We drew energy values after a transient of 3000 iterations of the recursion formula, which applies to every further computation. Different realizations of the chaotic sequence are generated by initiating the recursion with different initial values.

We can distinguish different ranges of values for the parameter $z$: periodic region, for $z<1$, in which the sequence converges to one fix point; and chaotic region for $z>1$, where $\epsilon_n$ varies over a fixed interval. Notice that while $z$ moves away from $z=1$ (value at which we obtain an ordered binary array) the $\epsilon_n$-values spread until this variable assumes almost all values in the interval $(0,1)$.

To compute the wave function for the tight binding model, we use the band matrix routines described in ref.\cite{Ting} using Gaussian elimination and back substitution.  Diagonalizing by this method one can handle large one dimensional systems reliably, in order to assess the localizing properties of the wave function. On the other hand, both the localization length and the conductance are more conveniently computed by renormalization procedure that generates the Green's Function $G(\epsilon)= [\epsilon + i\eta - H]^{-1}$ through a decimation method\cite{Ernesto}. This method consists in the elimination of sites along the chain in an iterative fashion, in such a way that we obtain an effective Hamiltonian which involves the sites 1 and $n+1$ in the chain, whose site energies are
\begin{eqnarray}
\tilde{\epsilon}_1(\epsilon) & = & \epsilon_1 + \Delta_{1(n)}^+(\epsilon),\\
\tilde{\epsilon}_{n+1}(\epsilon) & = & \epsilon_{n+1} + \Delta_{n+1}^-(\epsilon) .
\end{eqnarray}
Here $\Delta_{1(n)}^+(\epsilon)$ is the energy correction to the first site due the
presence of the other sites in the chain, and $\Delta_{n+1}^-(\epsilon)$ is the energy
correction to site $n+1$. These corrections are given in detail in ref. \cite{Ernesto,Lyra}.

In order to take into account the leads in the calculation procedure, we add a self energy correction to site energies $\tilde{\epsilon}_1$ and $\tilde{\epsilon}_{n+1}$\cite{Ernesto} whose value within the band $|\epsilon - \epsilon_0| \leq 2|V|$ is $\Sigma=\frac{\epsilon-\epsilon_0}{2} \pm i\sqrt{ V^2 -
(\frac{\epsilon-\epsilon_0}{2})^2  }$ where $\epsilon_0$ is the site energy of the semi-infinite chain representing the incoming and outgoing leads. Once the Green's function is calculated, the localization length $\xi$ is given by
\begin{equation}
1/\xi = \lim_{N\to\infty} \frac{1}{N} \ln{\left|\frac{G_{N,N}}{G_{1,N}} \right|} \; ,
\end{equation}
and the conductance by the two-lead Landauer formula, expressing the conductance in terms of the Green's function\cite{D'Amato},
\begin{equation}
g = \frac{2e^2}{h} \left ( 2\Gamma(\epsilon)|G_{1,N}|^22\Gamma(\epsilon)\right ),
\label{FisherLee}
\end{equation}
$\Gamma(\epsilon)$ being the imaginary part of the self-energy correction described above.

\begin{figure}
\includegraphics[width=7.7cm]{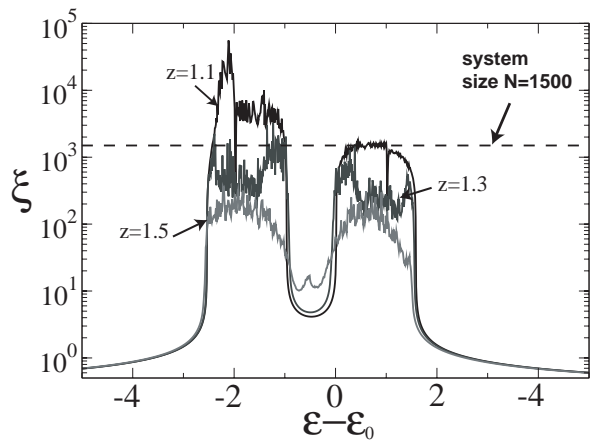}
\caption{\label{fig:length} The localization length as function of the energy for the shown values of $z$, $\epsilon_0=1$ and the coupling between sites set always to $V=-1$. The horizontal dashed line indicates the size of the system. Two bands of states appear in correspondance with the ordered alternating sequence close to $z=1^{+}$.}
\end{figure}

In Fig.\ref{fig:length} we plot the localization length as a function of the energy (referred to the site energies of the semi-infinite leads)  for three different values of the parameter $z$ in the vicinity of $z=1.1$. The dashed line shows the size of the system $N$ demonstrating that $\xi>N$ within a finite energy range. In this case two relevant  energy ranges are observed due to the nature of the initially binary energy sequence generated (see Fig.\ref{figure1}). The localization length grows linearly with the size of the system in the two  energy ranges shown. 

\begin{figure}
\includegraphics[width=7.9cm]{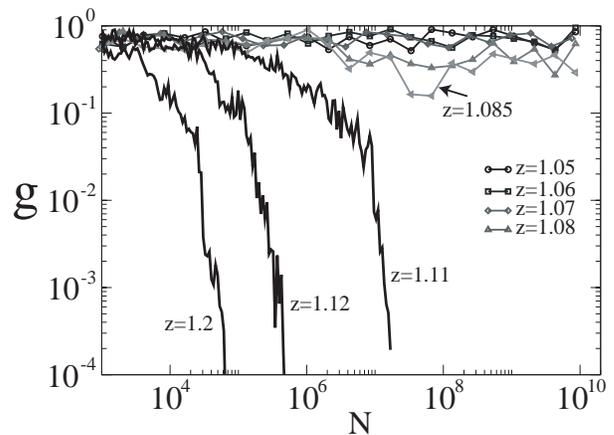}
\caption{\label{fig:conductance} Conductance $g$ (units $2e^2/h$) eversus size of the system for different values of $z$. The curves show a threshold value in the vicinity of $z=1.085$ above which the conductance decays exponentially. Each curve is averaged over $5$ different initializing seeds for the recursion generating chaotic sequences.}
\end{figure}

The delocalized nature of states within the energy bands observed in Fig.\ref{fig:length} can also be corroborated by computing the conductance according to Eq.\ref{FisherLee}. Fig.\ref{fig:conductance} shows the conductance $g$ as a function of system size for different values of $z$. Each curve is built by averaging over a few different initializing seeds for the chaotic sequence, avoiding spurious transient effects as described above. Below $z=1.085$, the conductance is essentially constant up to system sizes of $N=10^{10}$. Above the mentioned threshold value, the conductance drops exponentially towards zero, indicating a clearcut transition.

\begin{figure}[b]
\includegraphics[width=3.2in]{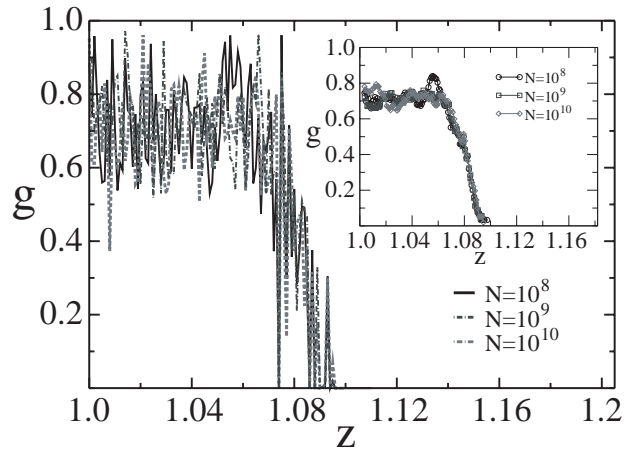}
\caption{\label{fig:phase5}Conductance (units $2e^2/h$) versus $z$ for linear chains of the sizes shown. The collapse of the curves demonstrates that no further changes operate on the curves and an asymptotic limit has been reached. The inset shows the running average of the curves that reduces noise and indicates a clearcut metal-insulator transition.}
\end{figure}

As a check that we have reached asymptotic behavior we have plotted the conductance $g$ as a function of the parameter $z$, increasing size until the curves reached a fixed limit. These results are shown in Fig.\ref{fig:phase5}, where no further changes are evident as we increase system sizes above $10^8$ sites. The figure suggests a clearcut transition between extended states and localized states. The inset shows a running average of the main curves so as to decrease fluctuations. The computation of the wavefunction versus the position along the one dimensional chain, as explained above Eq.\ref{eq:TB} led to the same conclusion. We diagonalized systems of up to $1000$ sites for different values of the parameter $z$ and energy $\epsilon-\epsilon_0=-1.5$. The wavefunction decays exponentially beyond a certain value of the control parameter. For $z=1.5$ the wavefunction is completely localized below 500 sites.

\begin{figure}
\includegraphics[width=9cm]{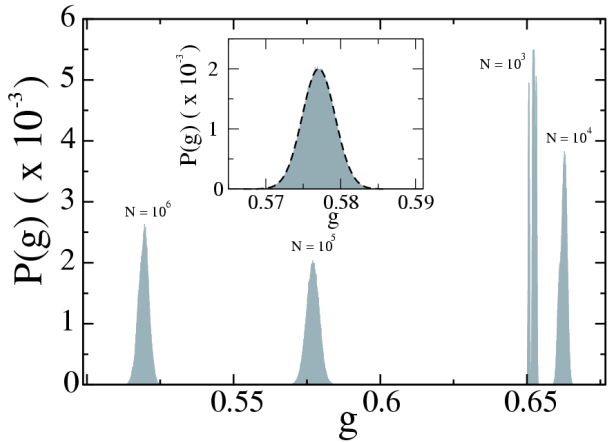}
\caption{\label{fig:distribution} Conductance distribution in the metallic phase as a function of systems size. Each distribution is generated by changing the seed of the chaotic energy sequence, and they are well defined Gaussians (see inset). Nevertheless changes in length involve larger fluctuations in the mean values.}
\end{figure}

To show that the average conductance is a good measure of the conductance distribution, the full distribution of the conductance is obtained as a function of system size and initializing seeds. In Fig.\ref{fig:distribution} we show that the distribution, when sampled for different initializing seeds, is a narrow Gaussian (see inset) of width 2\% of the mean conductance value. On the other hand, as the system size is changed the whole distribution fluctuates within a broader interval of 30\% around the mean conductance of Fig.\ref{fig:phase5}. These two different fluctuational amplitudes imply a non-selfaveraging property of the extended phase of the system rather different from a regular metallic phase. The distribution in the critical region, on the other hand is broad and requires a much more extensive study in itself.

As the results show, we obtain a metal-insulator transition for a finite band of states in a one dimensional system with a chaotic sequence of energies. The metallic character is assessed from the behavior of the localization length with system size, the scaling of the conductance and the behavior of the wavefunction as a function of the chaoticity parameter $z$. Nevertheless, for the energy sequence used, correlations are present but are not long ranged, so such feature seems sufficient but not necessary for delocalization to occur in the present sense.  One might be tempted to test a model where energies around $0$ and $1$ are pertubed randomly in the vinicity of those energies, generating a small band of values (resembling structure of Fig.\ref{figure1}. Nevertheless, such a model behaves asymptotically localized for arbitrarily small disorder.  We have thoroughly tested this conclusion with the methods described before.

Recently, several physical systems have been reported to be very sensitive to changes in the nature of disorder within the system. Among these systems we can mention: the two-dimensional Ising model, the problem of ballistic deposition, and random walks \cite{Ferrenberg, Souza, Gonzalez}. Regarding random walks, it is possible to show that if there is ``determinism'' in the external noise applied to random walks, the resulting walks observe symmetries which are absent when the noise is completely random. Anomalous behavior ensues in the computation of the mean distance versus time.

In the context of ballistic deposition, that serves as a prototype for studies of dynamic scaling phenomena in nonequilibrium growth processes, the external influences are  usually modeled by random noise. D' Souza {\it et al} have studied ballistic deposition using different deterministic sources of noise \cite{Souza}, and find fluctuations statistically inconsistent with the steady-state distribution and ergodicity i.e. driving the dynamics of the system with different deterministic sources of  ``noise'' results in selective sampling of the phase space. Such deterministic noise has been shown to modify the scaling exponents predicted by theory assuming strictly random noise.

The class of energy sequences addressed here seem to open new possibilities, which rather than sequence correlation, emphasize determinism implicit in chaos. Such determinism seems to play a non-trivial role in combination with the quantum mechanics of localization. This is a new example of physical systems where deterministic disorder can produce new phenomena. However, within quantum mechanics the phenomenon is more subtle: If the Lyapunov exponent that defines the chaos of the energy sequences is larger than certain critical value, the system behaves as the usual Anderson model.

\section{Acknowledgments}
We acknowledge useful discussions with A. Hasmy,  R. Paredes and L. S. Froufe P\'erez.

\newpage

\end{document}